\documentclass[3p,times]{elsarticle}

\usepackage{ecrc}

\volume{00}

\firstpage{1}

\journalname{Nuclear Physics A}

\runauth{Lokesh Kumar} 

\jid{nupha}

\jnltitlelogo{Nuclear Physics A}

\usepackage{graphicx}
\usepackage{amsmath,amssymb}

\usepackage{lineno}

\begin{document}

\begin{frontmatter}



\title{Systematics of Kinetic Freeze-out Properties in High Energy Collisions from STAR}

\author{Lokesh Kumar (for the STAR\fnref{col1} Collaboration)}
\fntext[col1] {A list of members of the STAR Collaboration and acknowledgements can be found at the end of this issue.}
\address{School of Physical Sciences, \\ National Institute of Science
Education and Research, \\ Bhubaneswar, India - 751005}




\begin{abstract}
The main aim of the RHIC Beam Energy Scan (BES) program is to
 explore the QCD phase diagram which includes search for a possible
 QCD critical point and the phase boundary between QGP and hadronic phase.
 We report the collision energy and centrality dependence of kinetic
 freeze-out properties from the measured mid-rapidity ($|y|<0.1)$ light hadrons
 (pions, kaons, protons and their anti-particles) for Au+Au collisions
 at the center-of-mass energy $\sqrt{s_{NN}} =$ 7.7, 11.5, 19.6, 27, and
 39 GeV. The STAR detector, with a large uniform acceptance and
 excellent particle identification is used in the data collection and analysis.
 The kinetic freeze-out temperature $T_{\rm{kin}}$ and average collective
 velocity $\langle \beta \rangle$ parameters are extracted from blast-wave fits
 to the identified hadron spectra and systematically compared with
 the results from other collision energies including those at AGS,
 SPS and LHC. It is found that all results fall into an anti-correlation
 band in the 2-dimension ($T_{\rm{kin}}$, $\langle \beta \rangle$) distribution: the largest
 value of collective velocity and lowest temperature is reached in
 the most central collisions at the highest collision energy. The
 energy dependence of these freeze-out parameters are discussed.

\end{abstract}

\begin{keyword}
Freeze-out \sep transverse momentum spectra  \sep beam energy scan

\end{keyword}

\end{frontmatter}



\section{Introduction}
\label{intro}
The ultra-relativistic heavy-ion collisions are expected to produce a
hot and dense form of matter called Quark Gluon Plasma (QGP)~\cite{qgp}. The
fireball produced in these collisions thermalizes rapidly leading to
expansion and cooling of the system. Subsequently, 
the hadronization takes place and the particles get detected
in the detectors. During this process, two
important stages occur as described below. The point in time after the collisions
when the inelastic interactions among the particles stop is referred
to as chemical freeze-out. 
The yields of most of the produced particles get fixed at
chemical freeze-out.
The statistical thermal models have successfully described the chemical
freeze-out stage with unique system parameters such as chemical
freeze-out temperature $T_{\rm{ch}}$ and baryon chemical potential
$\mu_B$~\cite{qgp,Andronic:2008ev,Wheaton:2004qb}. 
Even after the chemical freeze-out, the elastic interactions among the particles are
still ongoing which could lead to change in the momentum of
the particles. After some time, when the inter-particle
distance becomes so large that the elastic interactions stop, the
system is said to have undergone kinetic freeze-out. At this stage, the
transverse momentum $p_T$ spectra of the produced particles get
fixed. The hydrodynamics inspired models such as the Blast Wave Model~\cite{qgp,Schnedermann:1993ws,starpid}
have described the kinetic freeze-out scenario with a common temperature
$T_{\rm{kin}}$ and average transverse radial flow velocity $\langle \beta
\rangle$ which reflects the expansion in transverse direction. 

Quantum Chromo Dynamics (QCD), a theory of strong interactions,
suggests that the phase diagram has two main
phases: QGP and hadron gas. Lattice QCD
predicts that the transition between hadron gas and QGP is a
crossover~\cite{lattice1} at $\mu_B \sim$ 0. At
high $\mu_B$, the transition is expected to be a first
order~\cite{ejiri,kapusta}. In between, one expects the position of the
critical point, where the first order phase transition line ends~\cite{cp,Gupta:2011wh}. The 
experiments at RHIC focus on exploring the QCD
phase diagram, locating a critical point and determining the phase boundary between
hadron and QGP phase. In view of these, a Beam Energy Scan (BES) program was
started at RHIC~\cite{Kumar:2009eb,mohanty_npa,9gev,ref_bes}. The first phase of the BES program yielded many
interesting results as a function of energy or
$\mu_B$
related to
the search for 
critical point and phase boundary ~\cite{Kumar:2012fb,Kumar:2013cqa}. The kinetic freeze-out parameters
provide important information about the collision dynamics. The vast range of data
collected in the BES program allows for the systematic study of
kinetic freeze-out parameters and to see their energy dependence
trends. The BES results presented here are obtained for Au+Au
collisions at $\sqrt{s_{NN}}=$~7.7, 11.5, 19.6, 27, and 39 GeV at
mid-rapidity ($|y|<$0.1) using both STAR Time Projection Chamber (TPC)
and Time Of Flight (TOF) detectors~\cite{tpc_tof}. The error bars
shown in figures
represent the statistical and systematic errors added in
quadrature. The pion spectra presented here are corrected for the weak decay feed-down
and muon contamination while proton and anti-proton spectra are not
corrected for feed-down effects.

As mentioned earlier the kinetic freeze-out parameters $T_{\rm{kin}}$ and $\langle \beta \rangle$
of the system can be obtained using the hydrodynamics-motivated blast
wave model.
The model assumes that the particles are locally
thermalized at kinetic freeze-out temperature and are moving with a
common transverse collective flow velocity~\cite{Schnedermann:1993ws,starpid}. Assuming a radially boosted
thermal source, with a kinetic freeze-out temperature $T_{\rm{kin}}$
and a transverse radial flow velocity $\beta$, the transverse momentum
$p_T$ distribution of the particles can be given by 
\begin{equation}
\frac{dN}{p_T \, dp_T}  \propto  \int_0^R r \, dr \, m_T
I_0\left(\frac{p_T \sinh\rho(r)}{T_{\rm{kin}}}\right) 
\times K_1\left(\frac{m_T \cosh\rho(r)}{T_{\rm{kin}}}\right),
\end{equation}
where $m_T=\sqrt{p_T^2 + m^2}$, $m$ being mass of a hadron, $\rho(r)=\rm{tanh}^{-1}\beta$, and $I_0$ and
  $K_1$ are the modified Bessel functions. We use the flow velocity
  profile of the form $\beta=\beta_S(r/R)^n$, where $\beta_S$ is the
  surface velocity, $r/R$ is the relative radial position in the
  thermal source, and $n$ is the exponent of flow velocity profile. Average transverse radial flow velocity $\langle
  \beta \rangle$ can then be obtained as: $\langle \beta \rangle =
  \frac{2}{2+n} \beta_S$.

\begin{figure}
\begin{center}
\includegraphics*[width=9.cm]{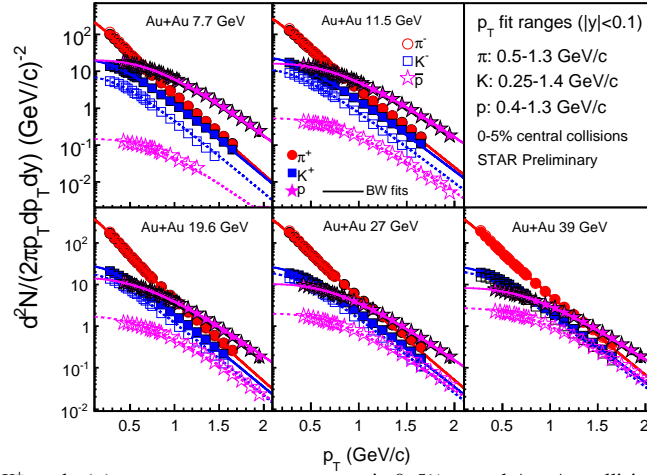}
\vspace{-0.9cm}
\caption{Invariant yields of $\pi^{\pm}$, $K^{\pm}$, and
      $p (\bar{p})$ versus transverse momentum in 0--5\% central Au+Au collisions at
      $\sqrt{s_{NN}}=$ 7.7, 11.5, 19.6, 27, and 39 GeV. 
Curves      represent blast wave model fits. 
}
\vspace{-0.5cm}
\label{fig:pt_spectra}
\end{center}
\end{figure}
\begin{figure}[htb]
\begin{center}
\includegraphics*[width=12.cm]{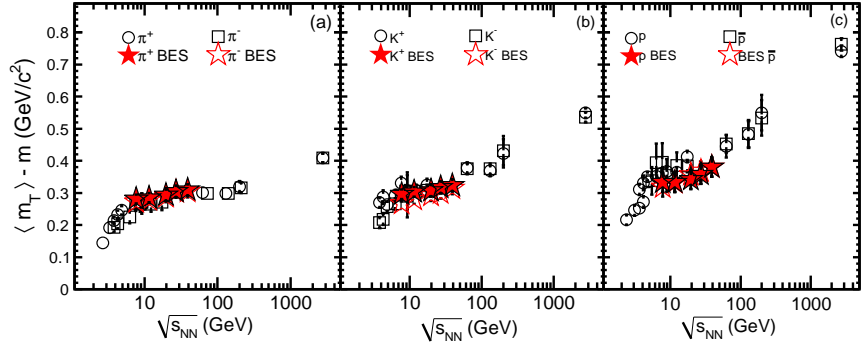}
\vspace{-0.9cm}
\caption{Energy dependence of $\langle m_T
\rangle-m$ for
 (a) $\pi^{\pm}$, (b) $K^{\pm}$,  and (c)  $p (\bar{p})$ in central
 heavy-ion collisions.
}
\label{fig:mt}
\end{center}
\end{figure}
Figure~\ref{fig:pt_spectra} shows the invariant yields of $\pi^{\pm}$, $K^{\pm}$, and
      $p (\bar{p})$ versus $p_T$ for $|y|<$ 0.1 in 0--5\% central Au+Au collisions at
      $\sqrt{s_{NN}}=$ 7.7, 11.5, 19.6, 27, and 39 GeV. 
These distributions
      are fitted with the blast wave model which can be seen
to reasonably describe the $p_T$ spectra of $\pi, K, p$ at
all energies studied. The low $p_T$ part of pion spectra is affected by resonance decays due
to which the pion spectra is fitted above $p_T>$~0.5~GeV/$c$. The fit
parameters are $T_{\rm{kin}}$, $\langle  \beta \rangle$, and $n$. 

\begin{figure}[htb]
\begin{center}
\includegraphics*[width=5.5cm]{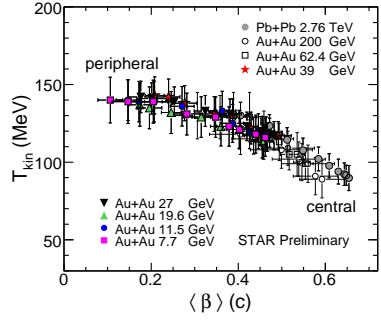}
\caption{
Variation of $T_{\rm{kin}}$ with $\langle  \beta \rangle$
for different energies and centralities. The centrality increases from
left to right for a given energy. The data points other than BES
energies are taken from Refs.~\cite{starpid,Abelev:2013vea}.}
\label{fig:tkin_beta}
\vspace{-0.6cm}
\end{center}
\end{figure}
The $p_T$ spectra can also be characterized by obtaining $\langle p_T \rangle$ or
$\langle m_T \rangle$, where $m_T$ is the transverse mass of the particles. Figure~\ref{fig:mt} shows the energy dependence of $\langle m_T
\rangle-m$ for $\pi^{\pm}$, $K^{\pm}$,  and  $p (\bar{p})$. The star symbols show
results from the BES while the data points for AGS, SPS, top RHIC, and
LHC energies (represented by open circles and squares), 
are taken from the
references~\cite{starpid,sps,ags,Abelev:2013vea}. It can be seen that $\langle m_T  \rangle-m$ increases with
energy at lower energies, remains almost constant at the BES energies
and then increases again towards higher energies up to the LHC. If the
system is assumed to be in a thermodynamic equilibrium, $\langle m_T  \rangle-m$ can be
related to temperature and $\sqrt{s_{NN}}$ can be related to entropy of
the system ($dN/dy$ $\propto$ $log(\sqrt{s_{NN}}$). In view of this,
the constant value of $\langle m_T  \rangle-m$ can be interpreted as a
signature of first order phase transition~\cite{vanhove}. However,
more studies may be needed to understand this behavior~\cite{bedanga_mt}. 

Figure~\ref{fig:tkin_beta} shows the variation of $T_{\rm{kin}}$ with $\langle  \beta \rangle$
for different energies and centralities. $T_{\rm{kin}}$ increases
from central to peripheral collisions suggesting a longer lived
fireball in central collisions, while $\langle  \beta \rangle$
decreases from central to peripheral collisions suggesting more rapid
expansion in central collisions. Furthermore, we observe
that these parameters show a two-dimensional
anti-correlation band. Higher values of $T_{\rm{kin}}$ correspond to
lower values of $\langle  \beta \rangle$ and vice-versa. 
\begin{figure}
\begin{center}
\includegraphics*[width=5.cm]{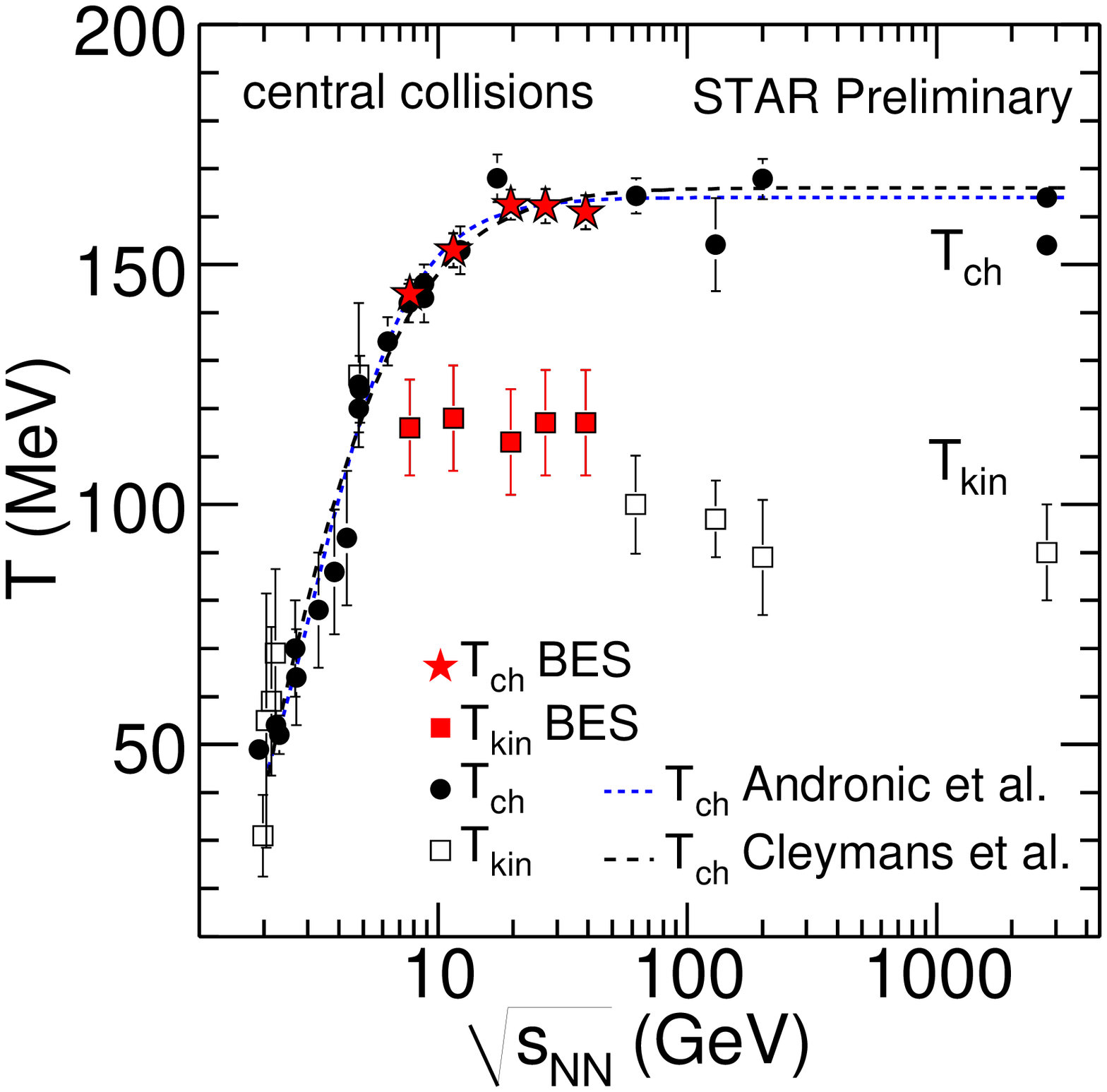}
\includegraphics*[width=5.cm]{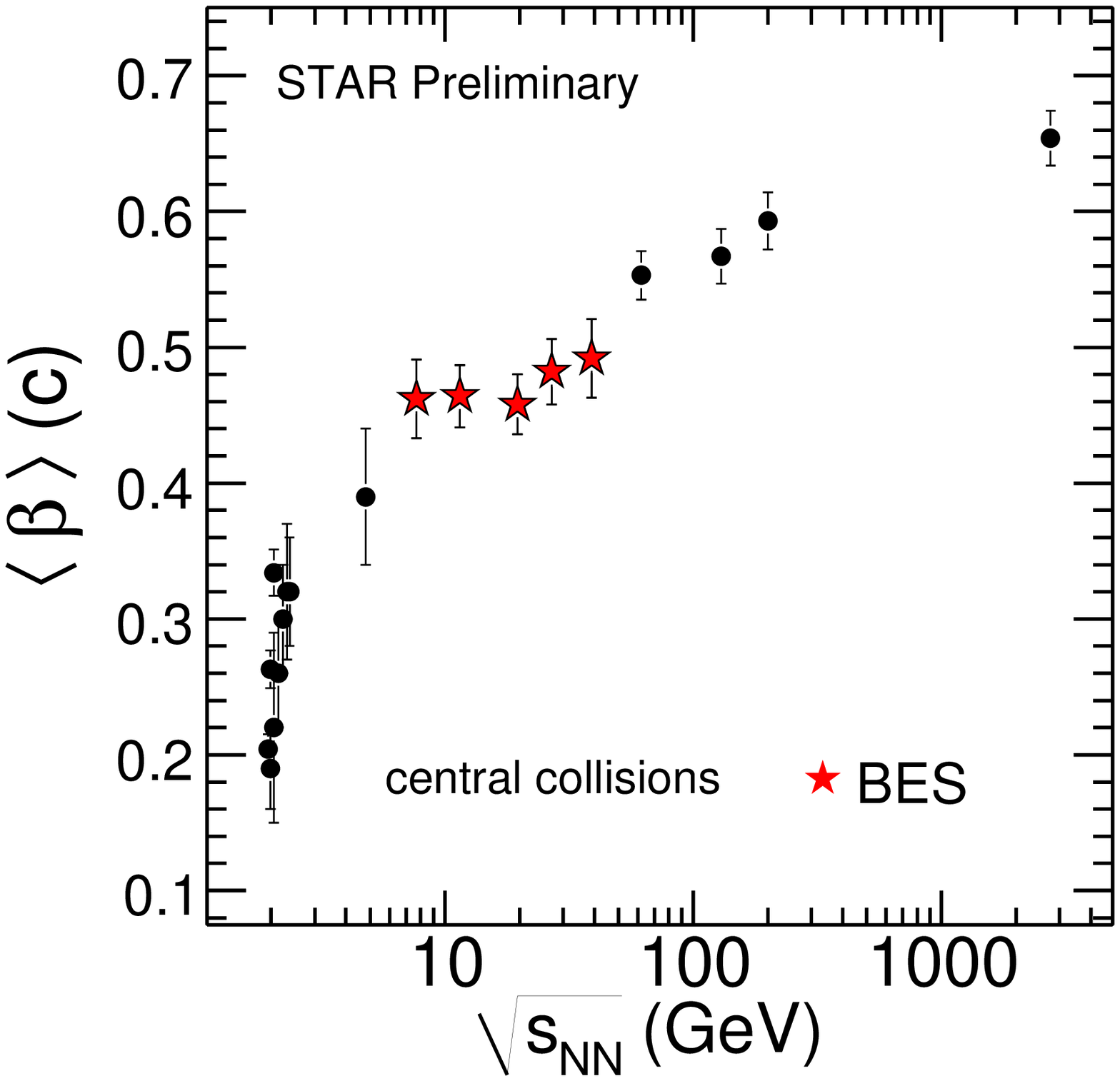}
\vspace{-0.4cm}
\caption{
Left panel: Energy dependence of kinetic and
chemical freeze-out temperatures for central heavy-ion collisions. The curves represent various
theoretical predictions~\cite{Cleymans:2005xv,Andronic:2009jd}.
Right panel: Energy dependence of average transverse radial flow
velocity for central heavy-ion collisions. The data points other than BES
energies are taken from
the Refs.~\cite{starpid,sps,ags,Abelev:2013vea} and references therein.
}
\vspace{-0.4cm}
\label{fig:tb_en}
\end{center}
\end{figure}

Figure~\ref{fig:tb_en} (left panel) shows the energy dependence of kinetic and
chemical freeze-out temperatures for central heavy-ion collisions. We observe that the values of kinetic and chemical
freeze-out temperatures are similar around $\sqrt{s_{NN}}=$4-5 GeV. If
the collision energy is increased, the chemical freeze-out temperature
increases and becomes constant after the $\sqrt{s_{NN}}=$11.5
GeV. However, the $T_{\rm{kin}}$ is almost constant around the
7.7--39 GeV and then decreases up to the LHC
energies. The separation between $T_{\rm{ch}}$ and  $T_{\rm{kin}}$
increases with increasing energy. This might
suggest the effect of increasing hadronic interactions between
chemical and kinetic freeze-out when we go towards higher energies~\cite{qgp}.
Figure~\ref{fig:tb_en} (right panel) shows the average transverse radial flow velocity plotted as a
function of  $\sqrt{s_{NN}}$. The $\langle \beta \rangle$ increases
very rapidly at lower energies, remains almost constant for
$\sqrt{s_{NN}}$=7.7--19.6 GeV, and then increases again up to the LHC
energies. 
Since $\langle \beta \rangle$ reflects the expansion in the
transverse direction, 
it is observed that this expansion velocity is
constant around $\sqrt{s_{NN}}$=7.7--19.6~GeV.

\section{Conclusions}
We have presented a systematic study of kinetic freeze-out parameters
in heavy-ion collisions with results from the RHIC BES program.
The $T_{\rm{kin}}$ increases from central to peripheral collisions suggesting a longer lived
fireball in central collisions, while $\langle  \beta \rangle$
decreases from central to peripheral collisions suggesting more rapid
expansion in central collisions.
The kinetic freeze-out temperature suggests a decrease from lower
($\sqrt{s_{NN}}\sim$ 4--5 GeV) to 
higher ($\sqrt{s_{NN}}\sim$ 2.76 TeV) energies. 
The separation between chemical and kinetic freeze-out
temperatures increases while going towards higher energies, suggesting
that hadronic interactions increase between chemical and kinetic
freeze-out at higher energies. 
The average transverse radial flow velocity increases rapidly at lower
energies ($\sqrt{s_{NN}}\sim$ 2--3 GeV) . 
The expansion in the radial direction remains similar around
$\sqrt{s_{NN}}=$7.7--19.6 GeV and then increases towards higher
energies where the initial energy
density produced is higher. The energy dependence of $\langle m_T
\rangle-m$ for $\pi$, $K$, $p$, and  $\bar{p}$ also shows similar constant
behavior around lower BES energies which could be related to
first-order phase transition signature. 








\end{document}